# The mysteries of fermions

Richard M. Weiner

Laboratoire de Physique Théorique, Univ. Paris-Sud, Orsay, France and

Physics Department, University of Marburg, Germany

Postal address: 112 Avenue Felix Faure, 75015 Paris

email: weineratstaffdotuni-marburgdotde

It is conjectured that all known fermions are topological solitons. This could explain the non-observation of bosonic leptons and baryons and provide a physical mechanism for the Pauli exclusion principle.

A most remarkable and so far unexplained observation in particle physics is that (A) all known fermions have baryon or lepton quantum numbers and that no bosonic leptons or baryons have yet been found [1].

Another specific property of fermions is (B) the Pauli exclusion principle, which states that, unlike in the case of bosons, a given quantum state cannot be occupied by more than one fermion. Although this principle is essentially a straightforward mathematical consequence of the spin-statistics theorem, the physics behind this property is not understood [2].

So far (A) and (B) were considered unrelated. It is the purpose of this paper to propose an explanation for (A) based on the Skyrme mechanism and which is consistent with the standard model; we will further argue that this also allows an understanding of (B). The possible simultaneous elucidation of two apparently independent mysteries of particle physics is an argument in favour of this mechanism and establishes a link between the two observations.

An explanation why spin ½ nucleons have baryon quantum numbers was suggested by Skyrme [3, 4] in terms of a quaternion-valued scalar field

$$U_{Skyrme} = \exp[(2i/f)\,\boldsymbol{\tau}\cdot\boldsymbol{\pi} \tag{1}$$

the Lagrangian [5] of which is non-linear:

$$\mathcal{L}_{Skyrme} = (1/16)\,f^2\,\mathrm{Tr}\,(\partial_\mu U_{Skyrme}\partial^\mu U_{Skyrme}^\dagger) +$$
$$+ (1/32e^2)\,\mathrm{Tr}\,[(\partial_\mu U_{Skyrme})\,U_{Skyrme}^\dagger,\,(\partial_\nu U_{Skyrme})\,U_{Skyrme}^\dagger]^2. \tag{2}$$

Here f is the pion decay constant, e a dimensionless parameter, $\boldsymbol{\pi}$ the pseudo-scalar pion triplet field and $\boldsymbol{\tau}$ represents the isospin Pauli matrices.

Skyrme proved that under certain conditions such a field can lead to topological solitons, i. e. states with degenerate vacua. These solitons may have half-integer spin and carry a topological quantum number, which can be identified with the baryon number.

The Skyrme approach to baryons got support [6] from the phenomenological success of the non-linear sigma model to which it reduces in a certain limit and from the link between the Skyrme model and QCD established by Witten [7]. Using the $N_{colour}$ expansion due to t'Hooft [8] Witten proved, among other things, that for odd $N_{colour}$, skyrmions actually have half-integer spin and that ordinary baryons can be understood as solitons in current algebra effective Lagrangians. New evidence for the soliton character of the baryon may have been found with the possible discovery [9] of excited baryons made of five quarks, which had been predicted by the QCD version of the sigma model.

It is natural to ask whether the Skyrme model, which may be considered as a bare-bone model for baryons in the sense that besides spin and baryon number it accounts qualitatively also for certain low-energy properties of strong interactions, cannot be applied to the analogous problem of leptons as well. Indeed, according to the standard model in the electroweak sector there also exists a non-linearly interacting scalar field, the Higgs field, which formally satisfies an equation similar to (1) (cf. below). Moreover, in analogy to the Skyrme field, which accounts for the spontaneously broken $SU(2)_L \times SU(2)_R$ symmetry of the non-linear sigma model, the Higgs field breaks spontaneously the corresponding $SU(2) \times U(1)$ symmetry. Actually, the similarity between the Higgs and the Skyrme field was observed a long time ago, although not in the present context. It is known that if the Higgs mass exceeds a certain limit of the order of $1/\sqrt{G_F} \approx 300$ GeV, where $G_F$ is the Fermi constant, at energies of this order weak interactions become strong and the $SU(2) \times U(1)$ electroweak sector of the standard model can also be approximated by a gauged non-linear sigma model [10]. Based on this result, Gipson and Tze [11] found that the electroweak sector, too, admits topological solitons, which could behave, what concerns spin and quantum numbers, as leptons. This observation, however, does not constitute the solution of our problem, because in the Gipson-Tze approach leptons are very heavy and interact strongly. More to the point is the paper by Tie-zhong Li [12] who considered the more general case of a Higgs field represented by two doublets rather than one, as postulated in the minimal standard model and who showed that in this case the analogy between the Skyrme model and the electroweak sector applies also for light and weakly interacting leptons.

To see this we start with the slightly generalized expression for the Skyrme Lagrangian [13], which holds also for non-vanishing pion masses $m_\pi$ when the chiral symmetry is broken

$$\mathcal{L}_{Skyrme} = (1/16) \, f^2 \, \text{Tr} \, (\partial_\mu U_{Skyrme} \partial^\mu U_{Skyrme}{}^\dagger) +$$

$$+ (1/32e^2) \, \text{Tr} \, [(\partial_\mu U_{Skyrme}) \, U_{Skyrme}{}^\dagger, (\partial_\nu U_{Skyrme}) \, U_{Skyrme}{}^\dagger]^2 +$$

$$+ (1/8) \, m_\pi^2 \, f^2 \, (\text{Tr} \, U - 2) \,. \tag{3}$$

Using (1) Eq. (3) can be written as an expansion in terms of the pion field **π**

$$\mathcal{L}_{Skyrme} = (1/2) \, \partial_\mu \boldsymbol{\pi} \cdot \partial^\mu \boldsymbol{\pi} - (1/2) \, m_\pi^2 \, \boldsymbol{\pi} \cdot \boldsymbol{\pi} + O(\boldsymbol{\pi}^4) \tag{4}$$

On the other hand if the Higgs consists of two doublets then there exists among the surviving five physical Higgs particles a pseudo-scalar triplet **H** the Lagrangian of which reads [12]

$$\mathcal{L}_{Higgs} = (1/2) \, \partial_\mu \mathbf{H} \cdot \partial^\mu \mathbf{H} - (1/2) \, m_H^2 \, \mathbf{H} \cdot \mathbf{H} + O(\mathbf{H}^4) \tag{5}$$

where $m_H$ is the Higgs mass. Expressing **H** in terms of a quaternion field

$$U_{Higgs} = \exp[(2i/F) \, \boldsymbol{\tau} \cdot \mathbf{H}] \tag{6}$$

the formal similarity between Eqs. (1) and (6) and Eqs. (4) and (5) respectively shows that the Higgs Lagrangian can be written as

$$\mathcal{L}_{Higgs} = (1/16) \, F^2 \, \text{Tr} \, (\partial_\mu U_{Higgs} \partial^\mu U_{Higgs}{}^\dagger) +$$

$$+ (1/32E^2) \, \text{Tr} \, [(\partial_\mu U_{Higgs}) \, U_{Higgs}{}^\dagger, (\partial_\nu U_{Higgs}) \, U_{Higgs}{}^\dagger]^2$$

$$+ (1/8) \, m_H^2 F^2 (\text{Tr} \, U - 2) \,. \tag{7}$$

In (7) F is a dimensional constant and E a dimensionless parameter.

In the limit of small masses $m_H$ the last term in Eq. (7) can be neglected and we are left with

$$\mathcal{L}_{\text{Higgs}} = (1/16) \, F^2 \, \text{Tr} \, (\partial_\mu U_{\text{Higgs}} \, \partial^\mu U_{\text{Higgs}}{}^\dagger) +$$

$$+ (1/32E^2) \, \text{Tr} \, [(\partial_\mu U_{\text{Higgs}}) \, U_{\text{Higgs}}{}^\dagger, (\partial_\nu U_{\text{Higgs}}) \, U_{\text{Higgs}}{}^\dagger]^2 \quad\quad (8)$$

which is the equivalent of Eq.(2).

We conclude that the analogy between the Skyrme Lagrangian and the Higgs Lagrangian holds also for small Higgs masses, provided the strong interaction constants f and e are replaced by corresponding electroweak constants F and E respectively, and the pionic pseodoscalar pion field **π** by the Higgs triplet **H**. Moreover, as shown by Adkins and Nappi [13], there also exist soliton solutions of Eq. (4) for certain non-vanishing values of $m_H$. Since for small values of $m_H$ the Higgs field is weakly interacting and since the formation of a soliton does not depend on the strength of the interaction, the Skyrme approach would apply also to weakly interacting leptons. Furthermore, because of the lepton-quark symmetry of the standard model one could conjecture that all known fermions are topological solitons and that the absence of bosonic leptons and baryons proves that at present energies this is the only mechanism nature chooses to produce fermions. It is this proposition which we now suggest, and discuss the consequences it leads to.

As a preamble we note that the possibility that the Higgs field is represented by a triplet is not only compatible with present data but has also some heuristic advantages as compared with the minimal standard model. Actually, triplet Higgs have become quite popular through the Little Higgs models [14]. These models provide also a direct approach to our weakly interacting lepton (quark) problem, since in these models there are light Higgs particles, which descend from non-linear sigma model fields that interact weakly up to energies of the order of 10-100 TeV. For these sigma model fields the Skyrme mechanism goes through as in the Gipson-Tze approach and one ends up again with skyrmionic leptons and quarks. These facts suggest that the Skyrme mechanism may be responsible for the observation that all known fermions have baryonic or leptonic quantum numbers.

If solitons are indeed the source of baryon and lepton quantum number and spin, one might possibly expect that there should not exist massless leptons and quarks, because (at a classical level, at least) solitons are extended, massive objects. Here from would follow that 1) *only particles without these quantum number can be massless;* 2) *even the lightest neutrino has non-vanishing mass.* The recent evidence for neutrino oscillations, which implies that at least some of the neutrinos have mass, fits nicely into this picture. As to the effective size of the leptons the following remark might be appropriate: if one identifies the Lagrangian (3) with that of the Higgs sector of the standard model (this might be appealing but is not necessary) the effective size is given essentially by the cut-off ~ FE, where the electroweak symmetry breaks down. Then the present experimental limits on lepton and quark compositeness would suggest that this cut-off is beyond 1 TeV, which would again favour Little Higgs models.

If all known fermions are indeed to be identified with skyrmions it is natural to ask whether the Skyrme mechanism is not also responsible for the exclusion principle, which is, besides the half-integer spin and the conserved quantum numbers, the main characteristic of fermions. We shall argue below that this indeed may be the case.

The Pauli principle for identical fermions manifests itself in two ways: in bound states, where it was also discovered and in scattering reactions. The existence of bound states of skyrmions that simulate quite satisfactorily light nuclei was proven among others in numerical simulations [15]. This confirms the action of the exclusion principle, without which nuclei could not exist.

In scattering processes the exclusion principle determines the short distance repulsion, which was already considered by Perring and Skyrme [4] (for a more recent discussion cf. Ref. [16]). In the hedgehog approximation for the pion field $\boldsymbol{\pi} = (\mathbf{r}/r)\,\theta(r)$ the potential V between two skyrmions at small distances r is positive and has a 1/r dependence. Moreover, as pointed out in Refs. [16,17] the repulsion is maximal for identical skyrmions. This is what one would expect if this repulsion effect is a manifestation of the Pauli principle.

But perhaps more relevant in the present context is the behaviour of high density skyrmion systems, since the statistical distribution of these systems is a sensitive function of the kind of statistics these systems satisfy.

In a study of the energy density ε of skyrmions at high number densities n Kutschera, Pethik and Ravenhall [17] derived from $\mathcal{L}_{\text{Skyrme}}$ the canonical result [18]

$$\varepsilon \sim n^{4/3} \qquad (9)$$

where the numerical coefficient in front of $n^{4/3}$ in Eq. (9) coincides within the uncertainties of the numerical calculation with that obtained for quark matter. However relation (9) is nothing else but the high density ultrarelativistic limit of the Fermi-Dirac distribution of fermions, which in its turn is a direct consequence of the Pauli exclusion principle.

Although (cf. also [16]) the Skyrme model for baryons is expected to work only at low energies, as pointed out by the authors of [17] the above result justifies aposteriori its use also at high densities, precisely because baryonic matter is expected to reduce at high densities to quark matter. Moreover, as will be shown below, the main dimensional argument of this derivation can also be used for the Higgs Lagrangian $\mathcal{L}_{Higgs}$ supposed to describe weakly interacting leptons and quarks and here this qualification does not apply. Indeed, consider U as defined in Eq. (1) of the form $U = U(\mathbf{r}/\lambda)$, where $\lambda$ represents a scale length. The space components of the current $J_\mu = (\partial_\mu U_{Skyrme}) U_{Skyrme}^\dagger$ scale as $\lambda^{-1}$, so that at high densities in the Lagrangian only the fourth order term $\sim \lambda^{-4}$ counts. Since the average particle density scales as $\lambda^{-3}$ the energy density has at high densities the form (9).

We conclude from these considerations that the same property of topological solitons which possibly explains the baryonic and leptonic quantum numbers of fermions may also explain the Pauli principle. In this way the exclusion principle loses much of its mysterious "fiat" character and becomes intuitively understandable.

A possible confirmation of this interpretation is represented by the Chandrasekhar limit, which amounts to the experimental fact that white dwarfs with masses above 1.4 solar masses have not been observed. This fact was predicted by Chandrasekhar [19] assuming, among other things, that in these stars the gravitational attraction energy overcomes the repulsive

Pauli energy of the degenerate electron gas and the stars collapse to black holes.

It is interesting to note that the particular value of 1.4 solar masses of the Chandrasekhar limit is a consequence of the particular form of the number density dependence of the energy density (9). In the derivation of Eq. (9) sketched above the $n^{4/3}$ dependence emerges as a consequence of the fourth order degree of the non-linear term in the Skyrme Lagrangian. The consistency of this result with that derived from the Fermi-Dirac distribution implies that $\mathcal{L}_{Higgs}$ does not contain higher than fourth order terms. Indeed it may be worth reminding here that the Chandrasekhar limit is quite sensitive to the exact power of n in (9). The value of 1.4 solar masses derived by Chandrasekhar [19] for the upper limit of white dwarfs is based on the 4/3 power of n, while the non-relativistic limit of Eq. (9) $\varepsilon \sim n^{5/3}$ assumed by Chandrasekhar's predecessors for the stellar electron gas leads to higher mass limits for white dwarfs, in contradiction with experiment. The scaling argument of Ref. [17] implies that sixth order terms in the Higgs Lagrangian would even lead to $\varepsilon \sim n^{6/3}$.

The suppression of the Pauli principle means that this principle is not absolute, as conventional quantum mechanics postulates, but is a dynamical property, which can be influenced and even annihilated by external attractive fields. The possibility of influencing and even suppressing the Pauli principle by adding a suitable term in the Lagrangian emerges naturally once this principle is understood as a consequence of the Skyrme Lagrangian. This relationship between the Pauli principle specific for fermions which have half-integer spin and which are usually associated with spinor fields, and the Skyrme field, which is a scalar, is natural once we have convinced ourselves that half-integer spin can emerge from such a non-linearly interacting scalar field.

The exclusion principle was initially *postulated* by Pauli to explain certain experimental facts. Much later it was derived within the spin-statistics theorem as a consequence, essentially, of the anti-commutation relations for fermions. The "derivation" of the Pauli principle from the Skyrme Lagrangian conjectured above implies therefore the "derivation" of the anti-commutation relations, which a priori look as mysterious as the Pauli principle.

The link suggested by the Skyrme mechanism between quantum numbers of fermions and the Pauli principle might imply that a possible violation the conservation of fermion quantum numbers is associated with the violation of the Pauli principle and viceversa.

We have conjectured that all known fermions are produced by the Skyrme mechanism and that this explains the non-observation of leptonic and baryonic bosons. The question arises whether the Skyrme mechanism remains the only mechanism nature chooses to create particles with baryonic or leptonic quantum numbers even beyond the present energy regime [20], i.e. that also beyond the present energy regime baryonic or leptonic bosons do not exist. This would contradict supersymmetry, which postulates that each fermion is associated with a boson with identical quantum numbers and that the non-observation of these particles at present energies is due to a breaking of this symmetry, the mechanism of which is yet unknown. The preceding considerations might imply that should bosonic leptons or baryons be observed at higher energies the Skyrme mechanism for fermions sketched above ceases to apply at these energies. This would establish a link between the breakdown of the Skyrme mechanism and the breaking of supersymmetry and possibly a clue for this last breaking .

We hope that the preceding considerations might contribute to the demystification of the properties of fermions. An important consequence of the identification of fermions with skyrmions is the existence of a non-linearly interacting effective scalar field both in the strong and electro-weak sectors. This consequence, which accounts for the universality of the Pauli principle, is quite general and expected to persist in possible generalizations of the standard model. For electroweak interactions this scalar field could, but has not necessarily to, coincide with the Higgs field. If scalar fields needed supplementary supportive arguments [21], the possible explanation of the Pauli principle and the non-observation of bosonic leptons and baryons might easily fulfil this task. That there are no bosons with fermionic quantum numbers but that the mechanism, which produces these quantum numbers (as well as the other characteristic property of fermions, half-integer spin) is possibly a boson field appears as an amusing paradox.

I am indebted to John Iliopoulos for reading an initial draft of this paper and for many valuable comments. I have also benefited from instructive discussions with Chia-Hsiung Tze in an earlier stage of this work. Last but not least the continuous help of Meinhard Mayer in clarifying the ideas of this paper and formulating them is most gratefully acknowledged.

was suggested by Finkelstein and Rubinstein (D. Finkelstein and J. Rubinstein, J. Math. Phys. **9,** 1762 (1968)  even before the formulation of supersymmetry and before the establishment of QCD, and without reference to the Higgs field. These authors, inspired by the Skyrme mechanism, proposed a model for fermions based on an analogy with rubber bands, which essentially anticipates topological solitons, and which made them conclude that "there are no truly neutral fermions" and that "no boson and fermion can have exactly the same set of values for their conserved particle numbers".

[21] For a lucid "defence" of scalars independent of the Skyrme mechanism cf. J. Iliopoulos, *Physics beyond the standard model*, Lectures presented at the 2007 European School of High Energy Physics, arXiv:0807.4841.